\def \AAP #1 #2 {{\em Astron. Astrophys.\/} {\bf #1}, #2}
\def \AAL #1 #2 {{\em Astron. Astrophys. Lett.\/} {\bf #1}, L#2}
\def \AAR #1 #2 {{\em Astron. Astrophys. Rev.\/} {\bf #1}, #2}
\def \AAS #1 #2 {{\em Astron. Astrophys. Suppl. Ser.\/} {\bf #1}, #2}
\def \AJ #1 #2 {{\em Astron. J.\/} {\bf #1}, #2}
\def \ANNREV #1 #2 {{\em Ann. Rev. Astron. Astrophys.\/} {\bf #1}, #2}
\def \APJ #1 #2 {{\em Astrophys. J.\/} {\bf #1}, #2}
\def \APJL #1 #2 {{\em Astrophys. J. Lett.\/} {\bf #1}, L#2}
\def \APJS #1 #2 {{\em Astrophys. J. Suppl.\/} {\bf #1}, #2}
\def \APSS #1 #2 {{\em Astrophys. Space Sci.\/} {\bf #1}, #2}
\def \ASR #1 #2 {{\em Adv. Space Res.\/} {\bf #1}, #2}
\def \BAIC #1 #2 {{\em Bull. Astron. Inst. Czechosl.\/} {\bf #1}, #2}
\def \JSQRT #1 #2 {{\em J. Quant. Spectrosc. Radiat. Transfer\/} {\bf #1}, #2}
\def \MN #1 #2 {{\em Mon. Not. R. Astr. Soc.\/} {\bf #1}, #2}
\def \MEM #1 #2 {{\em Mem. R. Astr. Soc.\/} {\bf #1}, #2}
\def \PLR #1 #2 {{\em Phys. Lett. Rev.\/} {\bf #1}, #2}
\def \PASJ #1 #2 {{\em Publ. Astron. Soc. Japan\/} {\bf #1}, #2}
\def \PASP #1 #2 {{\em Publ. Astr. Soc. Pacific\/} {\bf #1}, #2}
\def \NAT #1 #2 {{\em Nature\/} {\bf #1}, #2}
\def \SAIT #1 #2 {{\em Mem.\ Soc.\ Astron.\ It.\/} {\bf #1}, #2}
\def \MESS #1 #2 {{\em The Messenger\/} {\bf #1}, #2}
\def \ASTRNACH #1 #2 {{\em Astron. Nach.\/} {\bf #1}, #2}

\documentstyle[epsf,graphicx,rotating,lscape,amssymb]{BZpap}
\input epsf.sty
\begin{opening}
\title{The 2nd edition of the Roma-BZCAT}
\author{E. Massaro$^1$, P. Giommi$^2$, C. Leto$^2$, P. Marchegiani$^1$,\\
A. Maselli$^3$, M. Perri$^2$, S. Piranomonte$^4$, S. Sclavi$^1$}
\institute{$^1$Dipartimento di Fisica, Universit\`a di Roma ''La Sapienza'', Roma, Italy \\
$^2$ASI Science Data Center, ASI, Frascati, Italy \\
$^3$IFC, INAF, Palermo, Italy \\
$^4$Osservatorio Astronomico di Roma, INAF, Monteporzio Catone, Italy}
\date{} 
\end{opening}

\begin{document}

\oddpagefooter{}{}{} 
\evenpagefooter{}{}{} 
\medskip  

\begin{abstract} 
The 2nd edition of the Roma-BZCAT is available on line at the ASDC website 
({\sf http://www.asdc.asi.it/bzcat}) and in the NED database. 
In this short paper we describe the major updates from the first edition.

\end{abstract}

\medskip

\section{Introduction}
The {\it Roma-BZCAT} is a list of carefully checked Blazars originally conceived as a tool for 
the identification of counterparts of high energy sources.
The Fermi-LAT collaboration is in fact currently using it for this purpose in various $\gamma$-ray source 
catalogues, like 1FGL (Abdo et al. 2010a) and 1LAC (Fermi LAT AGN Catalog, Abdo et al. 2010b).

The complete first edition of the {\it Roma-BZCAT}, only available on-line included 2728 sources 
(Massaro et al. 2009). 
The first two volumes of a printed version covering the RA intervals 0h-6h (Volume I) and 6h-12h 
(Volume II) have also been published (Massaro et al. 2005, Massaro et al. 2008).
In addition to the catalogue tables, these two volumes present a large collection of 
data and the Spectral Energy Distributions (SED) of more than one hundred sources.

\section{The 2nd Edition }

The second edition of the {\it Roma-BZCAT} has been issued on line on May 11, 2010 (see 
{\sf http://www.asdc.asi.it/bzcat}).
Compared to the first edition it includes over 250 additional sources, many
of which discovered in very recent surveys.

Plotkin et al. (2010) presented a sample of 723 optically selected BL Lac 
candidates based on the Sloan Digital Sky Survey Data Release 7
(York et al. 2000). However, only 106 of these objects have been added so far to 
the {\it Roma-BZCAT} catalogue as we decided to adopt rather severe selection criteria.
In our first approach, we considered only sources with a radio flux
larger than 5 mJy at 1.4 GHz and preferentially with an optical spectrum
indicating a strong nuclear activity.
Fainter sources, in fact, could be safely distinguished from other types of
AGNs, like weak radio galaxies. 

Other new blazars and blazar candidates come from lists of possible associations 
with $\gamma$-ray sources discovered in the LAT survey (Abdo et al. 2010b).
New optical observations, most of which still unpublished, have been used to  
distinguish between BL Lac objects and FSRQs.
We accepted this classification, but we list the BL Lacs without an
available spectrum in the candidate section.

Other radio discovered blazar candidates, characterized by flat spectra, without
a firmly established optical counterpart, were not considered as
confirmed blazars and therefore are not included in this version of the catalogue.  

Fig. 1 (top) shows the sky distribution in equatorial coordinates of all sources 
reported in the 2nd edition of the {\it Roma-BZCAT}: as expected, the 
North-South asymmetry is more pronounced than in the 1st edition, because newly discovered 
blazars are mostly in the Northern sky.
The same asymmetry is apparent in the sky distribution in galactic coordinates, 
shown in Fig.1 (bottom), due to the absence of deep surveys in the Southern sky.

The $R$ magnitudes are taken from USNO databases and in a number of cases are affected by the presence 
either of large host galaxies or of very close sources.
In this case, magnitudes reported with the notes $g$ for a strong host galaxy contamination, 
$p$ and $t$ for pair and triple systems, do not correspond to the blazar
nuclear component. 
Moreover, radio fluxes at 1.4 GHz are mainly derived from NVSS (Condon et
a. 1998) and at 0.843 GHz from SUMMS (Mauch et al. 2003; indicated in
the catalogue by the note $s$) and can be affected by possible contributions 
originating from emission components different from the nuclear ones.
Finally, we recall that the completeness in flux level of various surveys at different 
frequencies used for discovering blazars and other AGNs is highly  non-uniform.

For these reasons it is very important to verify the selection strategy for
extracting samples to be used in statistical investigations to avoid the
occurrence of strong biases.

\section{Types of Blazars and source naming}
We use the same denomination of blazars adopted in the 1st edition. 
Each Blazar is identified by a three-letter code, where the first two are {\bf BZ}
for Blazar and the third one specifies the type, followed by the truncated 
equatorial coordinates (J2000).

The codes are:
\begin{itemize}
\item
{\bf BZB}: BL Lac objects, used for AGNs with a featureless optical spectrum, or
having only absorption lines of galaxian origin and weak and narrow emission lines;

\item
{\bf BZQ}: Flat Spectrum Radio Quasars, with an optical spectrum showing broad emission
lines and dominant Blazar characteristics;

\item
{\bf BZU}: Blazars of Uncertain type, adopted for a small number of sources having 
peculiar characteristics but also showing Blazar activity: for instance, occasional
presence/absence of broad spectral lines or features, transition objects between a
radio galaxy and a BL Lac, galaxies hosting a low luminosity Blazar Nucleus, ~~~etc.
\end{itemize}

The 2nd edition contains 1165 BZB sources, 261 of which are reported as candidates
because we could not find their optical spectra in the literature, 1660 BZQ sources
and 261 BZU objects.

\section{Scientific tools}
The on-line version of the {\it Roma-BZCAT} provides access to useful tools developed at ASDC that can be
easily accessed by clicking on the {\bf Data Explorer} button.
For instance it is possible to build sky maps of catalogued sources in the region surrounding the selected
blazar or retrieve optical and radio images at different size scale.
A large series of catalogues in many electromagnetic bands is available and 
all major databases can be accessed in a transparent way, including bibliographical
services.

An important tool is the {\bf ASDC SED builder}, that shows the Spectral Energy Distribution
of the source from a collection of available data.
SEDs can be enriched adding users data using the {\bf Load data} button.
It also possible to calculate some emission models by means of a Synchrotron-Self
Compton code.

Finally, the user can evaluate the spectral parameters useful to derive figures 
of merit of the sources for their high-energy detection according to different criteria.

\acknowledgements
We are grateful to the ASDC technical staff for the excellent work carried out to support
the on-line version of the {\it Roma-BZCAT}.
This work has been partially supported by research funds of Universit\`a di Roma
La Sapienza and by the ASI grant to INAF to support the scientific activities of the ASDC.

\clearpage

\begin{figure}
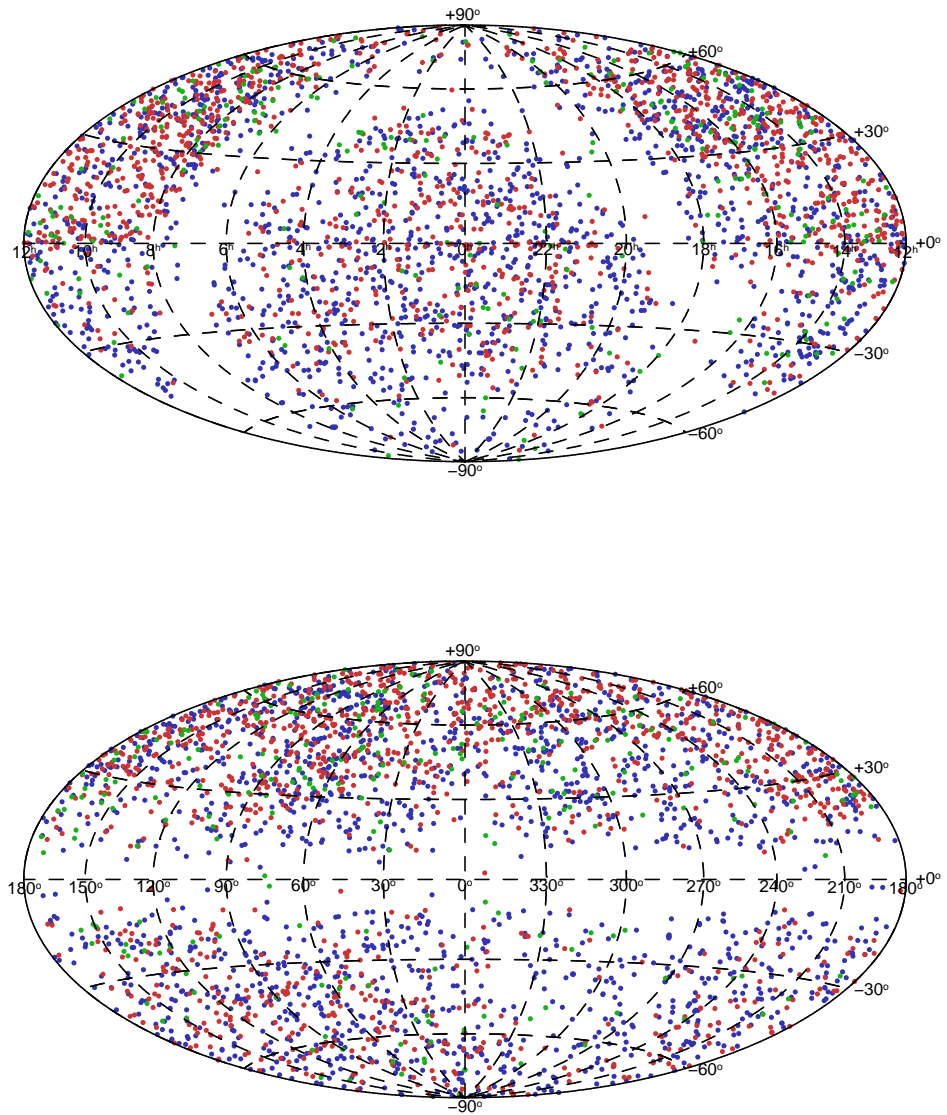


\includegraphics[height=15.5cm,angle=-90]{bzcatequatorial.ps} \\
\vspace{-2cm}
\includegraphics[height=15.5cm,angle=-90]{bzcatgal.ps}
\caption[h]{Hammer-Aitoff projection in equatorial (top) and Galactic (bottom) coordinates 
showing the distibution of Blazars included in the 2nd edition of the {\it Roma-BZCAT}. 
Red dots are BL Lacs and candidates, blue dots
are FSRQs and green dots are Blazars of uncertain classification.}
\end{figure}


\end{document}